# Importance of Secure Software Development Processes and Tools for Developers

1st Muhammad Danish Roshaidie
*School of Computer Science & Engineering,*
*Taylor's University*
Selangor, Malaysia
mddanishroshaidie@gmail.com

2nd William Pang Han Liang
*School of Computer Science & Engineering,*
*Taylor's University*
Selangor, Malaysia
panghanliang2@gmail.com

3rd Calvin Goh Kai Jun
*School of Computer Science & Engineering,*
*Taylor's University*
Selangor, Malaysia
goh98cal@gmail.com

4th Kok Hong Yew
*School of Computer Science & Engineering,*
*Taylor's University*
Selangor, Malaysia
hongyew99@gmail.com

5th Fatima-tuz-Zahra
*School of Computer Science & Engineering,*
*Taylor's University*
Selangor, Malaysia
fatemah.tuz.zahra@gmail.com

**Abstract** – In this research paper of secure software systems, authors have discussed what the proper development process is when it comes to creating a secure software, which will be suited for developers and relevent stakeholders alike. Secure Software Development Process for Developers is of crucial importance for software engineers as more and more software-based devices are becoming commonly available, and cloud services are evolving which require for the software to be constantly connected to the internet. With this in mind, Secure Software Development needs to be transformed to something most developers can rely upon to make applied software safe and have the capability to mitigate against potential attacks by hackers. Furthermore, in this paper, existing Secure Software Development Process ideas and implementations are reviewed and investigated using the research paper pool available online. Thereafter, an approach is proposed to enhance the security aspect in software development process to resolve security issues. Lastly, the paper concludes with final remarks on practical implementation of security features in software development phases for production of secure and reliable software programs and systems.

## 1.1 Introduction

Software application is becoming more and more commonly used in this digital world in addition to cloud computing services [1] and their enablement in various sectors like healthcare [2], transportation and many others. Software back then can be bought in shops where all the software you need is ready to be installed using a CD. The software is deemed to work as fine out of the box and has no updates or new features, until a new version of the particular software arrives, and till then you will need to buy that new CD. CD back then was a form of medium to get users to use and it can be easily abused and hacked, due to its lack of security updates. Software are also standalone and do not require any wireless systems to transmit data. Back then security vulnerabilities were not widespread as there was no internet connection needed to do so.

Nowadays, with the availability of the internet being widespread, installing software from a CD is likely old-fashioned where the user can just download the software easily from the browser. Software updates are becoming common where software companies want users to enjoy the new features without having to buy the software again. There is also a new business model when it comes to where Software As A Service (SAAS) is becoming common. It is a service where the user has to pay a licensing fee for a set of software from the software provider, receiving updates in a period of time. Once the software expires, the user should have to renew their license or they are no longer able to use that particular software anymore from the company. SAAS gives users a variety of functionalities that conventional software where they can get to access Cloud services, constant security updates and feature updates [3]. It gives users a piece of mind when it comes to security than the features it offers. Other software that also needs updates and constant security updates are autonomy software such as autonomous piloting of vehicles and the ever rising Internet Of Things [4]. In the report of the International Data Corporation (IDC),70% of companies have spent $1.2 billion on connectivity on 5G to be of use for IoT applications and business models [5]. These systems are also likely to be prone to attacks and need constant security updates and security measures.

There has been several incidents and news regarding hacking, especially the recent WannaCry ransomware (Fig. 1) which happened across the world, especially the hospital in the UK [6], in which their information was encrypted and unrecovered. The culprit is due to Windows ability to update and the hospital was running on an old firmware and current running on outdated software. The figure below was a screenshot of a ransomware pop up indicating that the ransomware software is running the background.

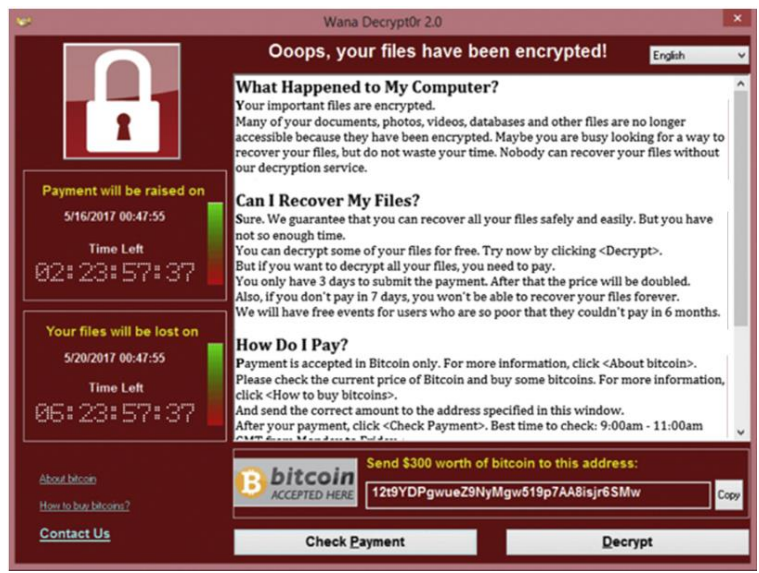

Fig. 1. Screenshot of WannaCry Ransomware [6]

When creating an application or a software what will be used by a lot of people, especially when it comes to making into an organisation, it is very logical to make sure the program or software is made to not only be easy to use and understandable to users, but also making sure the program itself be as secure as possible and be able to prevent from possible hacks or attacks that could jeopardize the whole program altogether. This is why during the development process, it is very crucial to create a software system implementation with security in mind to mitigate the possible risks mentioned [7].

It is a shame to have witnessed increase in number of security attacks on software in this age, and this is highly caused by the mentality among developers who think that software security is more of an afterthought, rather than something to be considered during the development process of the Software Development Life Cycle (SDLC) [8]. This problem could have been easily avoided by carefully planning security during the development process, and not that something to be thought of last minute or at the end of it. Every layer of software has to be as safe as possible from vulnerabilities that particular software is able to face. In addition to aforementioned cause of insecure applications there is another cause of increased security attacks. This is the evolution of internet, wireless networks and Internet of Things (IoT). Where life has apparently become convenient because of these technologies, they have also led to more insecurity in terms of data transmission, reliability of information and exposure to physical and cyber attacks [9]. For example, an increase has been observed in data breaches, network and routing attacks [10]. Therefore, it is important to implement security features in all phases of software development procedure as well as maintain and update the security components regularly. In next section the software development life cycle is reviewed to serve this purpose.

## 2    Literature Review

### 2.1    Secure Development Life Cycle

The Secure Development Life Cycle (SDL) is a process of fitting in security aspects in the typical Software Development Lifecycle (SDLC) (Fig. 2) used in many organizations developing software and systems [11]. There are many developers who follow the procedures of the SDL and apply them to a variety of other models such as Agile [12], Scrum, Waterfall and others. There are several other development models as well, like spiral, rapid application development, dynamic system development methodology, test driven development and others [13]. These models usually have similarities in terms of key phases involved which are: 1) planning, 2) design, 3) testing, 4) deployment, 5) maintenance, and 6) end of life.

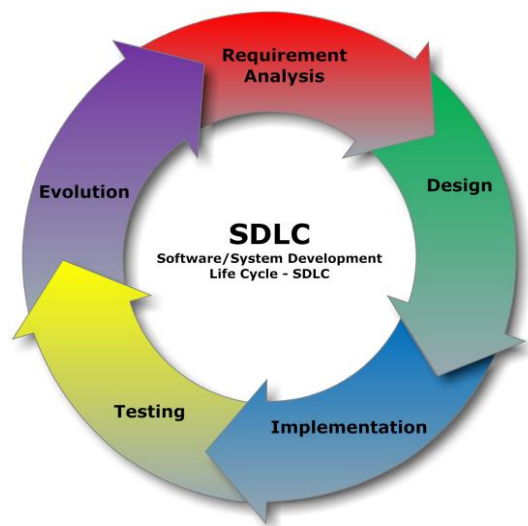

Fig. 2. Software Development Lifecycle [11]

Most of the security aspects that are commonly followed by developers happen in the testing phase, where developers test the software in place, finding the bugs and loopholes of the system [14]. This is one of the unhealthy practices when it comes to security since they are testing security of the software which does not have any form of security by design in the first place. This practice would result in time of deployment to be late and many resources will be spent on testing and improving the security instead. This is where having security to be implemented in SDLC as a concept is becoming more important that companies are widely adopting it, which is Secure SDLC. It is a development process where security-related features should be considered in all processes as possible.

In the investigation from [15], the development life cycle did not explicitly show the policy and guidelines that should be implemented in each management level during the Software Development Life Cycle (SDLC) phases. This scenario could lead to unpredicted threat and vulnerability embedded in the system [16]. In this case, SDLC should include Secure Development Lifecycle (SDL) to mitigate the risk of unknown threats and vulnerability. SDL promotes the idea of implementing security measures in every phase in the system development process [17]. Most software programmers develop program code solely to accomplish the requirement given by the client and not considering the security risk and vulnerability could possibly embed in the program that could be exploited by an attacker [18]. Malicious activities such as data disclosing, data alternating, denial of service could be presented at most by including security measures in the SDLC [19]. Figure 2 shows security measurement can be included in each phase of the SDLC.

From [20] research paper, every phase of SDLC should deliver functionality, absolute quality, and security measure. Although the aim of the software developer is to deliver all the functions in the shortest period. Casualty caused by security vulnerability might need more time, afford, and resources to resolve it. To build a secure software application you need to include security measures in the SDLC. The development process needs greater working time and does not guarantee 100% bulletproof to attack. But by applying the methodology minimizes the certainty of getting an attack. In the end, quality and secured software application. Thus, the reputation of the company will be gained from the public and peer groups.

Risk management is the security measure needed at the requirement analysis stage. It is a process of determining, accessing, and containing the threat that happens with the functions and Static analysis responsible for ensuring the system code is in proper coding convention and meets the standards of the field [18]. FindBugs from Bill Pugh and David Hovemeyer is a well known application used to perform static analysis for JAVA programming language. Another stage after the static analysis is security testing & code review. Secure code review has 8 components to focus on, which are authentication, authorization, session management data validation, error handling, logging, and encryption [21].

Security assessment and secure configuration are drawing all the steps to a close. It is a process to implement or update various types of security mechanisms to mitigate the threat and vulnerability found during the analysis phase [22]. The security mechanism for instance can be an encryption schemes [23] and protocols, authentication methods [24], user policies and regulations. Referring to [25], the human user is the root cause of system vulnerability. User security policy could regulate and control user behaviour as well as the authority to perform some activity [26].

## 2.2 Software Security Weaknesses

Developers in every stage have to know to be able to identify each of their weaknesses of security in their software development. There were plenty of terminologies based on a number of literature reports related to software security than can be defined for a better understanding, especially by Nabil Mohammad [14]. One of terminologies are the Asset, objects or items that are significantly important to the company and need to be kept safe. The assets are said to be the reason and main aims of a threat or attack to happen, and if those are exposed, it could face serious consequences for such data losses and thefts [27]. The software should be able to mitigate the attacks by hackers or make sure that it does not have too much reliance with data stored, or contained outside the assets.

Vulnerability in the software is an implementation within the security system that can potentially cause it to harm the software itself. It is a design flaw, or a loophole that can be found within the system itself. Software vulnerabilities are divided into two categories: one is at its design level, where the vulnerability itself is due to a design flaw, and two is at its implementation level, where the vulnerability is due to a bug that has been observed in the code. This however can be viewed as a threat to the system where the attacker has found the loophole, but hasn't compromised the assets. Another weakness that is relatable to vulnerability is the Defect of the secure software itself. This can be largely due to a design flaw of the security aspect of the software. However, unlike vulnerability, this flaw cannot be fixed and had to be entirely designed and replaced from the ground up. Defect should be detected as soon as possible so that when improving it, those defects have to be addressed and solved [28].

## 2.3 Problems in Secure Coding Practices

When it comes to the development stage of the cycle, it is at the design stage where it is crucial to ensure the inner workings of software development are assured to be safe and ready before actually developing the software. There are many cases where security is not a consideration in the design phase [7] in the first place which had led to the problem of security. Sometimes the security aspect is already had in mind, but not have taken it very seriously. There are a few key factors that could be the reason why there are bad practices happening around secure coding.

One which is quite prevalent on bad secure practices is the common miscommunication among developer teams. Sometimes, in the workplace no matter in which industries, there will be some slight misunderstanding where some teammates may have different views of how the software to be developed (for example). Furthermore, as the current situation wherever in the world where remote working is getting more important, most mentioned that working together physically is a better option [29] compared to working in a team. Communication is very crucial in this regard where developers and stakeholders should have the same understanding.

Another problem is the lack of software testing. This may be due to the bad practices of poor development scheduling which has taken up much resources in the Secure Software Development Life Cycle. By the lack of software testing, there will be potentially more zero-day-attacks once the software is deployed. This may be because of time constraints and effort required to perform bug vulnerabilities. However, in cases where testing is performed, it is appropriate to have a peer review of the software tests and not rely individually on testing matters just for the sake of time efficiency or to rush to meet the deadline. There have been surveys and reviews performed on this topic by researchers such as those in [30] might be helpful when considering to compare security features in recent past and present.

**2.4     Secure Coding Practices**

According to Veracode, "Security Skills are no longer an option for developers" [31] when it comes to software development, and it holds true given the frequent cyber security attacks occurred as mentioned above. Developers should follow on security development's best practices for that. There are 10 steps on approaching secure coding practices that should be followed by developers, no matter their background.

*1)   Check for Security Early and Frequently*

In a project, usually the security team will come at a later stage, which is at the end of the project before the deployment of the development, which should be a standard process of yesteryears. Having the security team last minute is not ideal, given by the time needed to test and go back to the drawing board to fix again. This has to be changed where security of data should be thought thoroughly in the beginning. Using OWASP Application Security Verification Standard as a guide is a welcome step in the right direction to help define the requirements of protections and gathering case conditions. There are certainly 10 most common OWASP vulnerabilities found as of 2020 by Sucuri in [32].

- Code Injection- where the hacker sends a code to the web application to make it perform actions it is not intended to do. An example of injection attack is SQL injection (Fig. 3).
- Broken Authentication- where the hacker use manual or automatic ways to gain a portion of access or control of the system, sometimes fully
- Sensitive Data Exposure- where sensitive information such as credentials, personal identification is compromised and not encrypted is seen by prying eyes of an attacker
- XML External Entities- a hack that was done by placing an XML input into the system, due to the poorly set up XML parser
- Broken Access Control- the access that wasn't meant to be granted is broken into
- Security Misconfigured- The flaw in the system itself due to lack of updates, unprotected files and unused inventories and features
- Cross- Site Scripting (XSS)- a classic vulnerability that is widespread and happens to plenty of websites, similar to code injection
- Insecure Deserialization- Encrypted data that is sent can be cracked and modified to gain access into the system, such as super cookie

- Vulnerable Applications- The system with unnecessary dependencies and components which are potential to be compromised
- Insufficient Logging and Monitoring- the lack of moderation of the system to detect for possible penetrations as soon as possible

Fig. 3. An example of SQL injection which could disclose and compromise sensitive data [32]

*2) Circling down Queries*

In case there is an attack being compromised due to a piece of code or a software, it is best for having practices of containing the piece of code by commenting out or removing it. At the same time developers have to be careful on allowing any inputs by the user when entering into sensitive object queries and advanced queries that the framework supports.

*3) Encoding the Data*

Encoding means to turn a raw string of input from a user into a harmless string that could be ineffective against such threats to the system. For example, if the input of the attacker is an injection script, the input is encrypted and performs like a normal string without having the piece of code that actually checks the input to think that that input is actually a piece of code as well.

*4) Validate All Inputs*

Sometimes you can't trust much of what the user might enter into the web application or system. There should be a set of rules set in place to make sure the syntax is valid, which is expected by the system. By having validation, you can narrow down smaller possibilities of the system to get attacked by hackers since some syntax are whitelisted, therefore acceptable and safe [33].

*5) Setting Up Identity and Authentication Controls*

Strong authentication methods are really important if you want to restrict users to have a number of controls to the system. By having controls with identity authentications in place, the program is less likely to have security breaches. Usernames and passwords are quite commonly used, so further enhancing the authentication method is much favourable in secure software implementations. There are plenty of authentication controls available, either by a one-time password, 2-factor authentication [34], or even biometric recognition [35] which is commonly seen in smartphones. If there is a user who happens to forget their password and resets to a new one, further security measures should be implemented as well to reduce chances of attackers gaining access and not the own user itself.

*6) Setting Up Access Controls*

Authentication controls shouldn't just stop there where users can simply gain access to every single feature of the system. Create a hierarchy of control access based on positions and level of users so that certain users have limited or more control to the system. If the position of the user is not a moderator level, that particular user should not have the ability to alter information or check logs of others, for example.

*7) Contain Data for Protection*

No matter what is the level of security of your system and the organization, it is a standard practice to keep sensitive data from prying eyes and away from the application as much as possible. Good practice is when the important data is encrypted while during transmission or currently idling in the application itself [36]. Cryptographic libraries are widely available and developers should not necessarily need to develop one in-house by themselves. The difficult aspects of cryptography such as key management, design, tier, and trust issues for a software that is complex should not be neglected in any way.

*8) Implement Logging and Intrusion Detection*

Relying on automated mitigation systems and extra authentication access isn't enough to justify its software security when there are no features of logging movements. Logs are crucial to check every movement done by users who have been in and out of the attack, and must be moderated at least by a user and not a robot. If an attack occurred, the logs can be useful to track for possible traces of where the origin of the attack happened, and be noted for improvements down the line. The software should log as much information as possible, such as timestamps, source IP, and user identification. For network hardware, there are many solutions that can be configured like an intrusion detection system [37][38], intrusion prevention system, or configuring honeypots and demilitarized zones (DMZ) [39][40].

*9) Utilising Security Frameworks and Libraries*

Developing security controls from scratch for any software or application can be a waste of resources, effort, and time, especially when the home-made software could create security flaws to it. There are plenty of well-established security frameworks that are available to be applied and could be better than creating from scratch. However, it is best to not just apply the security and call it a day. Adding additional securities into the system could significantly improve its overall security effectiveness, and frameworks used should be up to date and look out for any potential loopholes within.

*10) Look out for Errors and Exception Handling*

The final best practices to secure software is no less than monitoring and bug-testing. Errors and Exception Handling are examples of bad software engineering, especially in secure software development. Exceptions should be managed in a centralized manner to mitigate duplication of try/ catch blocks in the code. During the testing stage, error handling should be identified and reviewed carefully.

**2.5   Static Analysis-based Approach**

There was a study from a group of students who have been researching static analysis- based approaches when it comes to secure software development [8], especially when it comes to software vulnerabilities. The automatic static analysis is proved to be a very good security bug detector to be able to detect early bugs in the SDLC process. The analysis comprises a few key areas of security detection aspects, such as Vulnerable

Prediction Modelling, Software Metrics, and Static Analysis. With these areas covered, it can be possible to search for possible risks in the software itself which could take a fraction of the time, saving time and resources for development.

## 3   Methodology

There are various ways to collect results of the development for secured software. Collecting results of the problems and requirements of the software helps in understanding the appropriate secure software designs needed to be implemented and kind of asset that can be considered in development. All the results have been collected from research databases and sources such as IEEE, google scholar, research gate and ACM (Association for Computing Machinery). This method is called online research. Authors decided to use this method because all the research papers that are in the trusted online sources used existing data for their study. Using online research methods can be efficient and inexpensive because we are predominantly using research that has already been completed. By using this method, we can also access data across the world. One of the most important reasons to use this method is because most of the articles in journals contain graphs, charts, pictures and photos, images, etc., which helps in explaining the information that we are searching for visually and graphically.

Although the information is sourced from wide range of sources, accessing this information online is so much simpler because all we need is a computer and internet which is convenient in this Covid-19 pandemic situation. Narrowing the search to a focused angle makes it easier because all we need is to type in keywords that we are looking for and in just a few seconds we can get thousands of search results. Authors have focused on the current trends and best practices when it comes to the approach of secure coding practices. The reason why this is done is because software security is constantly improving and old practices are rendered very obsolete, even if the solution is a year old or two. Besides that, academic research papers contain real-life case studies which makes them excellent sources due to in-depth expertise and information. Academic research papers are proven to be never changed and properly moderated by a number of educated panels in the academic field of science. Lastly, several online scholarly journals come with a free download option that allows us to save material to our PC or smartphones. It helps us to access information even when there is no internet connection around. This helps save time by not having to wait for it to load for a long time.

## 4   Discussion

### A.   Best Practices of Secure Software [41]

Some interesting topics explored through literature in this paper are the different phases of SDL process that is easy to follow for developers from different IT backgrounds that acts as a guide model to help developers keep the security of their developing system as a top priority. The first phase of the Software Development Lifecycle (SDL) focuses on the concept of the system like what it is used for and what are the target users so that developers have a better defined security subsequently compliance to the project objectives. The first of phase of SDL is known as concept and planning which consist of multiple recommended practices, the first practise in this phase is SDL discovery that was mentioned and after the definition of systems security and compliance objectives, as this practise suggest choosing an SDL methodology to layout the plans that especially the plans that have interactions with SDL activities.

The SDL discovery practice is essential before any start of a project because it is able to help developers recognize security flaws during the development process with the intention that the development team can provide mitigation swiftly. Another well-known practice of the first phase of SDL is called security requirements as the name of the practice suggests is to list down potential security requirements regardless of front-end or back-end because this practice assists developers in determining possible issues and resolve them as soon as possible particularly in non-compliant areas of the developing system [42]. The last practise in this phase is known as security awareness training is essential to developers despite having experience this is due to technology is always updating and senior developers should be aware of any changes in the SDL models that generally improve their security and planning of a project [43]. Software development is a discipline which can either be centralized or decentralized. More recently distributed software engineering and component-based software development [44] has also become a trend with the advent and evolution of internet and online communication facilities which has simultaneously led to new types of challenges as well [45].

The next phase of SDL is also known as design and architecture that also consist of three practises such as threat modelling, secure design and tracking third party vulnerabilities. This phase focuses on the design of the product that meets the requirements with the practice of threat modelling that helps to identifying potential cyber attack plots with that in mind developers is able to provide remedies to the application in the current design phase by doing so it reduces the cost and time consume and consequently providing a incident response plan for future threats [46]. The next important practise of the second phase is secure design that emphasizes on the documentation of the design and its consecutive updates before being implemented in order to detect security risk so it can be reviewed and re-evaluated properly before it is available as a system to the public [47].

The last practise in this phase is third party software tracking also focus on security issues because these elements may potentially compromise the integrity of the system. Thus, regular maintenance and heavy monitoring of third party access is required so that patch updates can enhance and improvise the system security integrity. The third phase or stage in SDL is implementation where the project is carried out and the system is implemented with the help of three practices like secure coding, scanning tools, and code review. Secure coding in the third phase specifically targets the back-end of the project or the system and acts as a reminder for developers or debuggers to use terms that are encrypted in the source code of the system to reduce vulnerabilities of the system [48].

Meanwhile, scanning tools like Static Application Scanning Tools (SAST) are automatic scanners that scan through the source code to identify the security flaws that do not require the system to run it is recommended to scan on a daily basis from the start of the development of the project until it is implemented so that the risk can be identified thoroughly [49]. The last practise in this phase is code review while there are tools to help accelerate the process of code review like SASTs but it is still advised to regularly review the code manually after every scanning tool used to perform a double check for security issues that cannot be detected by tools.

The fourth phase of the SDL is the testing and bug fixing that contains three best practices, they are known as dynamic scanning also known as Dynamic Application Security Testing (DAST) and penetration testing. The first practise in this phase is dynamic scanning which involves the implemented system of the project while it is running. This an example of a running simulation of what cyber-criminals will exploit this can help developers have a better overview of the system through observing the configuration of the system's security before officially implementing it [50]. Penetration testing also known as Pentest is basically white hat hacking from other professional developers to test the integrity of the system. Into the bargain, the fuzzing test is also one the suggested practice for the fourth phase this practise help to enhance the protection of the system by becoming immune to attack types like SQL injection after processing it with a series of randomly generated input [51].

The fifth phase of SDL is deployment and maintenance that comprises two major practises. For instance, response plan and security checks. One of the topics of practise in this phase is incident response plan is a proposal steps of actions to take in any case of cyberattack and allows the monitoring team to quickly address issues faced in a short amount of time [52]. The second practice is the ongoing security checks which assist in maintaining the integrity of the system in the security aspects, this practice is recommended to carry out on a daily basis so that any vulnerabilities can be detected and the incident response team can react quickly with procedure plans [53].

The sixth phase of SDL is called the end of life support where developers no longer provide any updates to the implemented system and the data accumulated during the use of it time is caused to undergo proper disposals. There are two specific practises that handle the proper discarding of raw data if the system is to be terminated completely. The first practise for this last phase is data retention is the continuation of the storage of data despite not receiving updates but it is required due to companies retention policies for compliance and following the legal process of user's data protection to circumvent unnecessary lawsuits from clients or users [54]. The second practice for this phase is data disposal where a discard system is created specifically purging unwanted data with scrupulous attention to details and archived data must be protected with encryption in order to provide a secure method of retrieval for future uses [55].

## 5   Proposed Approach to Solve Security Issues

Based on the research, authors came up with a better solution to address the shortcomings of secure software development among developers. It is important to have solutions to address the problems so that every developer can learn and understand more about good secure software practices. One of the main solutions to improve the shortcomings of knowledge of developers about security is to require them to have more training and education in the area security. Possibly, the main problems addressed above are that the developers have little to no exposure towards security nor do they know much about the importance of security, therefore, education in this discipline would help them consider more on secure coding and applying methods of programming that could reduce the security risks.

Another solution proposed is to make developers utilize as many secure tools that are readily available and proven to be effective. As mentioned in the secure coding practices, developers may think that creating on their own would be a good security mitigating risk. However, it is a waste of time and not worth the effort if their own creation gets bypassed by attackers. Therefore, it is very encouraging for developers to utilize open source frameworks which have been proven to be effective in terms of security provision. If an organization wants developers to create a system or a web application with security as an integral component, they should consider their developers to always have security checks in all stages of the development process. Even when the project is deployed, there should be some level of security logging for the administrators and moderators to look for potential flaws in the system and detect attacks as early as possible to avoid larger harm later.

## 6   Conclusion

Undeniably, there are huge amount of software applications developed every day and from the human users at home to big corporations are highly dependent on the software applications to complete their daily tasks. In this case, ensuring the privacy, security and safety of each software application need to be the essential aims of each software developing body as well as software application reviewing commission. In fact, software developers did not pay much attention to securing software applications; instead they only delivered the minimum-security measure to run the software in the past and it has not changed much in the present as well. As we know that prevention is better than cure, therefore, existing research papers have been reviewed in this regard and it is observed including the cost of implementing security measures is far lesser than the cost used to mitigate an attack. An in-depth discussion on every phase of the Software Development Lifecycle and Secure Development Lifecycle is also carried out to provide the relevant stakeholders with information on security aspects of a software development process. A conclusion that could be made through the review and analysis performed in this paper is that security measures could be implemented in each phase of the software development lifecycle simultaneously. Some of them include risk assessment, designing threat models, code reviewing with the help of bug scanning tools, and system testing. Each security mechanism plays its role to find out the threat and vulnerabilities embedded in the system. Possible mitigation plan to respond to the worst-case scenarios that could possibly happen in the future have also been presented. Reflecting to the constantly changing technological environment, it could not guarantee zero vulnerability despite having every security measure. The proposed solution is to utilise security scanning tools and provide training to stakeholders in a frequent manner in addition to regular upgrade of the implemented software system.